\documentclass[12pt]{article}
\usepackage{epsf}

\newcommand{\bee}{\begin{equation}}
\newcommand{\ee}{\end{equation}}
\newcommand{\bea}{\begin{eqnarray}}
\newcommand{\eea}{\end{eqnarray}}
\newcommand{\R}{\rm I\kern-.2emR}
\newcommand{\C}{\rm \kern.25em\vrule height1.4ex
depth-.12ex width.06em\kern-.31em C}
\newcommand{\N}{{\rm I\kern-.16em N}}
\newcommand{\Z}{{\rm Z\kern-.35em Z}}

\begin{document}                                                                
\begin{flushright}
AZPH-TH-98-07\\
MPI-PhT/98-40\\
\end{flushright}
\bigskip\bigskip\begin{center}
{\Huge Nonlinear $\sigma$-model, form factors and universality
}
\end{center}
\centerline{Adrian Patrascioiu}
\centerline{\it Physics Department, University of Arizona,}
\centerline{\it Tucson, AZ 85721, U.S.A.}
\centerline{\it e-mail: patrasci@ccit.arizona.edu}
\vskip5mm
\centerline{and}
\centerline{Erhard Seiler}
\centerline{\it Max-Planck-Institut f\"ur Physik}
\centerline{\it (Werner-Heisenberg-Institut)}
\centerline{\it F\"ohringer Ring 6, 80805 Munich, Germany}
\centerline{\it e-mail:ehs@mppmu.mpg.de}
\bigskip \nopagebreak

\begin{abstract}
We report the results of a very high statistics Monte Carlo study of the
continuum limit of the two dimensional $O(3)$ non-linear $\sigma$ model.
We find a significant discrepancy between the continuum extrapolation
of our data and the form factor prediction of Balog and Niedermaier,
inspired by the Zamolodchikovs' $S$-matrix ansatz. On the other hand
our results for the $O(3)$ and the dodecahedron model are consistent 
with our earlier finding that the two models possess the same continuum 
limit.
\end{abstract}
\vskip2mm

In a recent paper \cite{PL} we reported some striking numerical results:
the continuum limit of the two dimensional ($2D$) $O(3)$ nonlinear 
$\sigma$ model seems to agree as well with the form factor prediction 
\cite{bn} inpired by Zamolodchikovs' $S$-matrix as with the continuum
limit of the dodecahedron spin model. The latter, known rigorously to
possess a phase transition at nonzero temperature, is almost certainly not
asymptotically free. Zamolodchikovs' ansatz on the other hand is believed
to embody asymptotic freedom, a belief which was reenforced by some 
remarkable scaling properties discovered by Balog and Niedermaier
\cite{bn2}.

Initially  \cite{bncom} we thought that the resolution of this
apparent paradox must come from the absence of asymptotic freedom in the
form factor approach and that the scaling proposed by Balog and 
Niedermaier must actually be false. Soon after though we realized that 
the trouble was much more serious: indeed from the direct computation of 
Balog and Niedermaier of the 2,4 and 6 particle contribution it followed 
that the asymptotic value of the transverse current two point function 
$J(p)$ must obey the inequality
\bee
                 J(\infty)>3.651
\ee
\cite{bnlong}. On the other hand in two recent papers \cite{conf1,conf2} 
we used reflection positivity to prove rigorously an upper bound on
$J(\infty)$ in terms of $\beta_{crt}$. In those papers the bound is given
only for the $O(2)$ model, but it is straightforward to generalize it
to any $O(N)$. For $O(3)$ it reads
\bee
                 J(\infty)\leq {2\over 3}\beta_{crt}
\ee
Combining these two inequalities one would conclude that $\beta_{crt}>5.5$.
This lower bound is so large that irrespective of the Balog and 
Niedermaier scaling ansatz, it suggests that the critical point may very 
well be at infinity and the model asymptotically free. 

Since, as stated above, asymptotic freedom of $O(3)$
is hard to reconcile with its having the same continuum limit as the
dodecahedron spin model, we decided that we needed a more refined
numerical study to decide these issues. In the present paper we will
report the results of our investigation regarding the continuum limit
of the $O(3)$ and dodecahedron spin models. We strove to achieve very
high numerical accuracy and to our knowledge our accuracy, exceeding
sometimes .05\%, has never been achieved before. Our results suggest
that in fact the model defined by the form factor approach does not agree
with the continuum limit of the $O(3)$ model. On the other hand, the 
numerical evidence is consistent with the hypothesis that the $O(3)$ and 
the dodecahedron spin models have the same continuum limit.

We begin our report by specifying the models and defining the
quantities measured. In both models the action is nearest neighbour,
the lattice square of length $L$, with periodic boundary conditions and
the inverse temperature $\beta$. The quantities measured are (with 
$P=(p,0)$):
\begin{itemize}
\item Spin 2-point function 
\bee
G(p)={1\over L^2}\langle |\hat s(P)|^2\rangle;\ \
\hat s(P)=\sum_x e^{iPx} s(x)
\ee
\item Current 2-point function $J(p)\equiv F(0)-F(p)$ with
\bee
F(p)={1\over 3 L^2}\sum_{a=1}^3  \langle |\hat j_2^a(P)|^2\rangle;\ \
\hat j_\mu^a(P)=\sum_x e^{iPx} j_\mu^a(x)
\ee
where \\$j_\mu^a(x)=\beta\epsilon_{abc}s_b(x)s_c(x+\hat\mu)$ \\
\item Magnetic susceptibility $\chi$:
\bee
\chi=G(0)
\ee
\item Effective correlation length $\xi$:
\bee
\xi={1\over 2\sin(\pi/L)}\sqrt{(G(0)/G(1)-1)}
\ee
\item Zero momentum renormalized coupling:
\bee
g_r=-{\bar u_4\over \chi^2\xi^2}
\ee
where $\bar u_4$ is the invariant connected spin 4-point function at zero
momentum.
\end{itemize}

Some comments regarding these quantities:
\begin{itemize}
\item Our definition of $\xi$ differs from that of Balog and Niedermaier,
who define $1/\xi^2$ as the position of the pole in the spin 2-point
function (on our Euclidean lattice, that pole is inaccessible).
If we implement our definition upon the form factor prediction, 
using $p\xi=1$ as the lowest nonvanishing momentum, we find
$\xi=1.0016187$, compared to 1.
\item The renormalized spin 2-point function is
$G_r(p\xi)=G(p\xi)/G(0)$
(normalized so that $G_r(0)=1$).
\item For the $O(3)$ model the definition of the current, including the
factor of $\beta$, is fixed by the
Ward identity. The quantity $J_{O(3)}(p\xi)$ is a `renormalization
group invariant', which describes continuum physics in the limit
$\xi\to\infty$, $p\xi$ fixed.
\item For the dodecahedron spin model the definition of the current is 
ad hoc, but if in fact its continuum limit agrees with that of $O(3)$ it 
is possible that upon multiplication by a suitable factor $c(\xi)$,
$c(\xi)J_{dod}(p\xi)$ reaches the same limit as $J_{O(3)}(p\xi)$.
The results reported in \cite{PL} seemed to corroborate this
assumption. 
\item Besides being a renormalization group invariant, $g_r$ vanishes in
any free field theory. Hence $g_r$ tests the nontrivial part of the
$S$-matrix.
\end{itemize}

The numerics were performed on a SGI-2000 machine. Both models were
simulated using the cluster algorithm and all quantities, except $g_r$,
measured using Wolff's improved estimator \cite{wolff}. At every
$\beta$ and $L$ value we performed runs consisting of 100,000
thermalization clusters and 1,000,000 clusters used for measurements.
These runs were repeated for $O(3)$ up to 149 times and for the
dodecahedron up to 246 times. We give the results for $\chi$ and
$\xi$ in Tab.1 ($O(3)$) and Tab.2 (dodecahedron).
As can be seen, our results for $O(3)$ are
more precise than for the dodecahedron, in spite of the extra
effort invested in the latter. The errors were estimated using the 
jack-knife method on the results of different runs, since this
led to a larger estimated error than the error estimated in each run.

\begin{figure}[htb]
\centerline{\epsfxsize=8.0cm\epsfbox{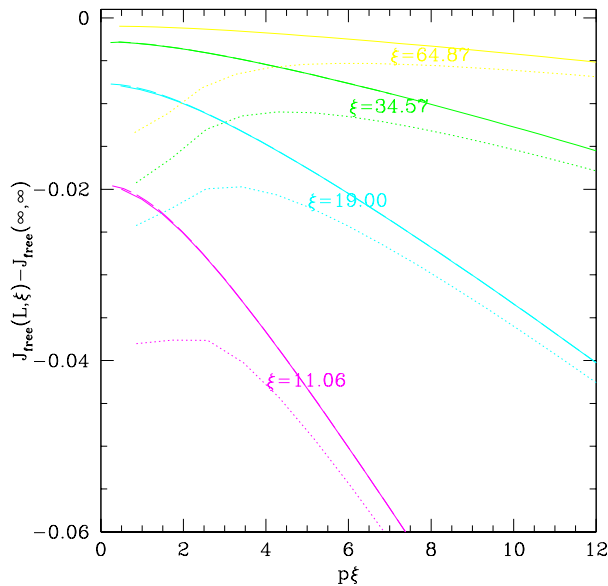}}
\caption{Lattice and finite volume artefacts for the complex free field;
dotted curves: $L/\xi\approx 7$, solid lines: $L/\xi\approx 14$, dashed
lines: $L/\xi\approx 24$ -- the latter two are practically 
indistinguishable.}
\label{freefield}
\end{figure}

The form factor prediction gives the continuum, infinite volume values
of $G_r(p\xi)$ and $J(p\xi)$ (however truncated as to the number of
intermediate particles included). To relate Monte Carlo measurements to
these predictions one must take both the continuum limit, $\xi\to\infty$,
and the thermodynamic limit, $L/\xi\to\infty$. In Fig.\ref{freefield}
we present the deviations from the continuum, infinite volume limit for 
the free field, displaying $J(p\xi)$ for several correlation lengths 
$\xi$ and ratios $L/\xi$. It can be seen that for $p\xi<12$ for 
$L/\xi\geq 13$ the finite volume corrections become negligible (less than
$2\times 10^{-4}$) for $\xi\geq 11$ (the smallest correlation length in 
our study). We verified directly that the same is true for the $O(3)$ 
model by running at the same $\xi$ values on $L/\xi$ approximately 7-8, 
13-15 and 23-25.

Another lesson learned from Fig.\ref{freefield} concerns the approach to 
the continuum: at fixed $p\xi$ the approach is described by
\bee
              {a/\xi^2}+b\log(\xi)/\xi^2
\ee
with $a$ and $b$ depending upon $p\xi$. Since at small $p\xi$ the $O(3)$
data are reasonably close to the free field values, we employed this
ansatz in all our extrapolations to the continuum limit.

\begin{figure}[htb]
\centerline{\epsfxsize=8.0cm\epsfbox{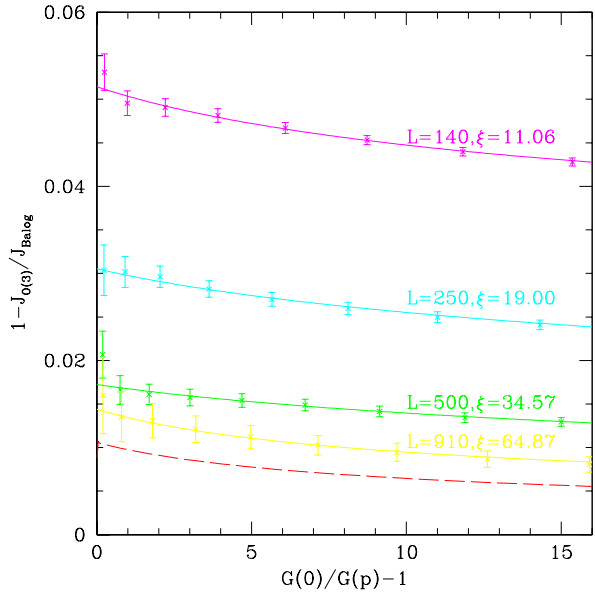}}
\caption{Comparison of Monte-Carlo ($J_{O(3)}$) with form factor
results ($J_{ff}$); the dashed line represents the continuum 
extrapolation.}
\label{o3bal}
\end{figure}

\begin{figure}[htb]
\centerline{\epsfxsize=8.0cm\epsfbox{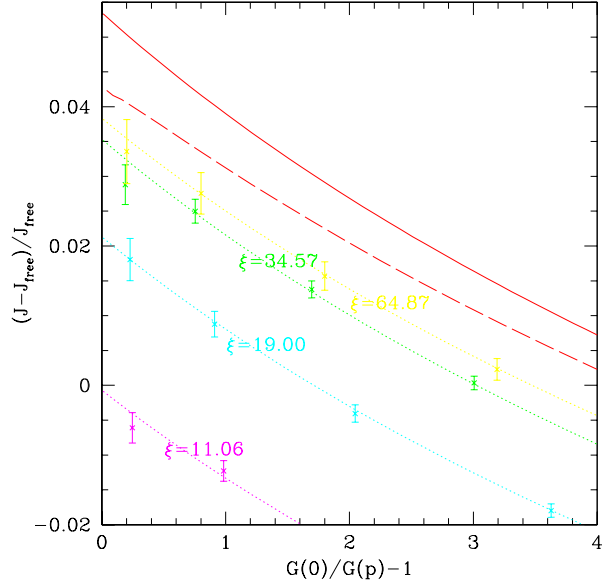}}
\caption{Comparison of Monte-Carlo ($J_{O(3)}$ -- dotted lines) and their
continuum extrapolation (dashed line) with form factor results ($J_{ff}$ 
-- solid line): interacting contribution.}
\label{o3balint}
\end{figure}

\begin{figure}[htb]
\centerline{\epsfxsize=8.0cm\epsfbox{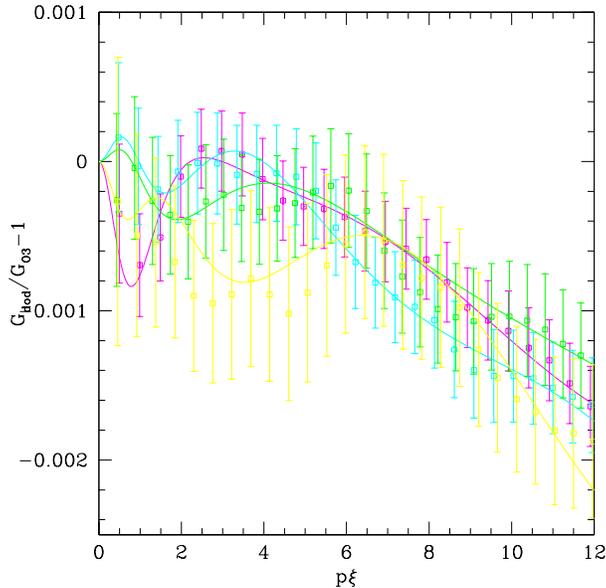}}
\caption{Comparison of $G_r$ for $O(3)$ and dodecahedron. Data with 
$\xi\approx 11,\ 19,\ 34,\ 65$ are overlaid. The solid lines are fits.
 }
\label{dod}
\end{figure}

\begin{figure}[htb]
\centerline{\epsfxsize=8.0cm\epsfbox{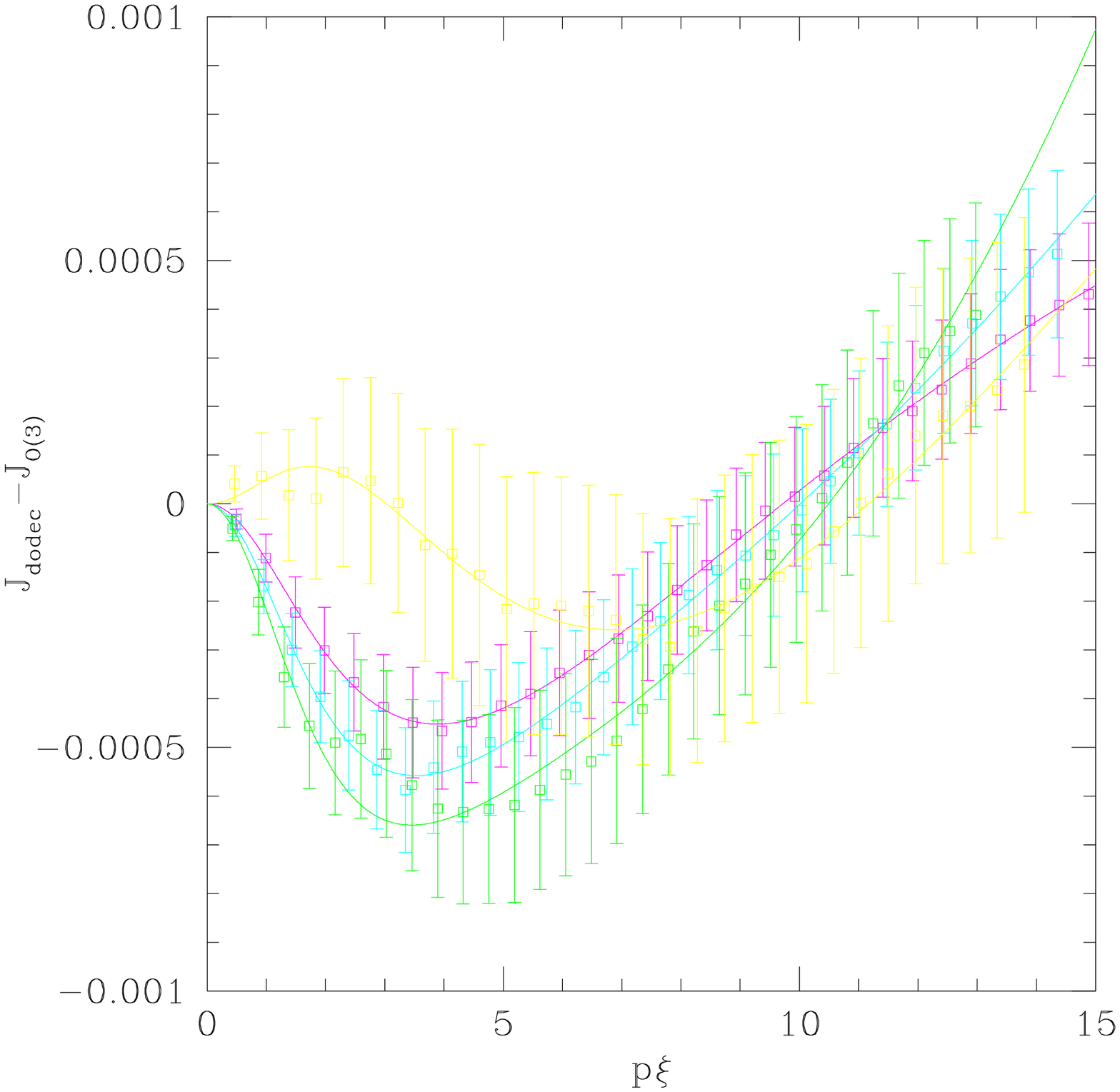}}
\caption{Comparison of $J$ for $O(3)$ and dodecahedron (renormalized).
Data with $\xi\approx 11,\ 19,\ 34,\ 65$ are overlaid. The solid lines 
are fits.
 }
\label{jdod}
\end{figure}

As stated above, our definition of $\xi$ differs from that of Balog and 
Niedermaier. Even though we could correct for this difference by applying
our definition to their data, we decided that a better approach is to 
eliminate $\xi$ altogether and plot $J$ versus $G(0)/G(p)$. For small 
$p\xi$ the latter quantity is very close to its free field value, 
$1+(p\xi)^2$. This plot is shown in Fig.\ref{o3bal} for various $\xi$ 
values and $L/\xi$ aproximately 14. The solid lines represent a least 
square fit to the data of the form
\bee
\sum_{k=1}^4 a_k (J_{free}(p\xi))^k
\ee
where 
\bee
J_{free}={1\over 2\pi}-{1\over\pi}\sqrt{1+z/4\over z}
\log\Biggl(\sqrt{1+{z\over 4}}+\sqrt{{z\over 4}}\Biggr)
\ee   
with $z=(p\xi)^2$ is the expression corresponding to $J$ in the
free two-component scalar field theory. We also show the extrapolation of
our results to the continuum limit using the ansatz in eq.(3) and the 
form factor prediction. As it can be seen, the latter disagrees with the 
former.

The disagreement is statistically significant, yet very small. One may 
wonder how such a wonderful agreement could occur if the form factor
prediction is actually false. A possible answer is this: the correlation
functions $J$ and $G_r$ are, especially at low momenta, very close to 
those of the free theory, deviating only by less than 5\%. The reason
behind this is that the form factor squares contributing to the spectral
functions of the correlation functions are at low momenta
\begin{itemize}
\item dominated by the lowest possible number of intermediate particles
(1 for $G_r$, 2 for $J$)
\item the squares of those lowest form factors are very close (in the case
of $G_r$ even equal) to those of the free theory.
\end{itemize}
Therefore, if we want to test the form factor prediction, this universal 
free field value should be subtracted. To achieve such a
subtraction we do the following: we plot $x^2J(x)$ versus $G(0)/G(x)$ 
$x=p\xi$ for both the $O(3)$ model and the free field. We then subtract 
the two curves at the same value of the abscissa and plot the difference.
This is shown in Fig.\ref{o3balint}, together with the form factor 
prediction. The latter differs from the continuum value extrapolation by 
about 25\% at $p=0$.

Another quantity which measures directly the nontrivial part of the 
$S$-matrix is the renormalized coupling $g_r$. For the $O(3)$ model this 
quantity was determined to be around 6.5--6.9 \cite{nc} and we determined
it again at $\beta=1.5$, $\xi=11.06$ to $g_r=6.565(25)$ and at $\beta=1.6$
to $g_r=6.609(18)$. Since $g_r$ is expected to be monotonically decreasing
in $\beta$, the small difference between the two values is probably just
due to statistical fluctuations. The value of $g_r$ in the form factor 
approach has not yet been calculated, even though the ingredients are 
the same as for $G(p)$. We think that the comparison would be most
interesting.

Next we would like to present the comparison of the dodecahedron spin 
model with the $O(3)$ model. If the two models possess the same continuum
limit, then one would expect that for sufficiently large $\xi$ both the
lattice artefacts and the finite volume corrections become identical.
Consequently if we use the same definition for $\xi$ in both models, it
should be legitimate to compare the same `renormalization group 
invariants' at the same $p\xi$ value. In Fig.\ref{dod} we present the 
comparison of $G_r$.
The data are consistent with the hypothesis that the two models have the 
same continuum limit. The error bars are dominated by the those coming 
from the dodecahedron, where the fluctuations are very large. We
present a similar comparison in Fig.\ref{dod} for $J$, where the 
renormalization was chosen so that the rms deviation between the two 
models in the range from $p\xi=0$ to $p\xi=15$ is minimized. As stated 
above, the theoretical status of this procedure is not clear. Finally we 
report on $g_r$ for the dodecahdron. We measured it at $\xi=11$ and 19 
and obtained 6.42(84) and 6.33(83). These values suggest that its 
continuum limit value is also around 6.5, in agreement with the value of 
$g_r$ for $O(3)$.

Our conclusion is that whereas there is good numerical evidence that the
model defined via the form factor approach does not agree with the
continuum limit of the lattice
$O(3)$ model, there is no evidence that this limit is different from that
of the dodecahedron model. These facts represent a major setback for the 
accepted saga regarding the special properties of the nonlinear $\sigma$
models with $N\geq 3$.

We acknowledge numerous discussions with J.Balog and
M.Nie\-der\-mai\-er. A.P. is grateful to the Max-Planck-Institut for
its hospitality.

{\bf Tab.1:}\\
{\it Susceptibility $\chi$ and effective correlation length
$\xi$ for $O(3)$}

\begin{tabular}[t]{r|r|r|r|r}
$\beta$ & $L$ &\# of runs & $\xi$& $\chi$ \\
\hline
\hline       1.5 & $140 $ & $149$ & 11.0600(29) & 176.7348(566) \\
\hline
1.6 & $250 $ & $86$  & 18.9971(55) & 448.3651(1515) \\
\hline
1.7 & $500 $ & $100$ & 34.5712(97) & 1270.8185(4196) \\
\hline
1.8 & $910 $ & $44$  & 64.8723(277)& 3839.8609(2.1498) \\
\end{tabular}\\

{\bf Tab.2:}\\
{\it Susceptibility~$\chi$ and effective correlation length
$\xi$ for the dodecahedron}

\begin{tabular}[t]{r|r|r|r|r}
$\beta$ & $L$ &\# of runs & $\xi$& $\chi$ \\
\hline
\hline
1.49   & $140 $ & $246$ & 11.0540(40) & 176.7967(840) \\
\hline
1.5835 & $250 $ & $201$  & 19.0337(55) & 451.1633(2320) \\
\hline
1.672  & $500 $ & $139$ & 34.4139(167) & 1265.2500(7350) \\
\hline
1.76   & $910 $ & $78$  & 66.6195(522)& 4046.1542(3.9885) \\
\end{tabular}\\

\end{document}